\documentclass[aps,prc,groupedaddress,twocolumn,showpacs,amsmath,amssymb]{revtex4}
\usepackage{graphics}% Include figure files
\usepackage{dcolumn}% Align table columns on decimal point 
\usepackage{longtable}

\begin{document}

\title{Alpha-particle optical potential proofs at astrophysically relevant energies}

\author{M.~Avrigeanu}
\author{V.~Avrigeanu}
\email{vavrig@ifin.nipne.ro}
%\homepage{http://tandem.nipne.ro/~vavrig}
\affiliation{"Horia Hulubei" National Institute for Physics and 
	Nuclear Engineering, P.O. Box MG-6, %077125 
	Bucharest--Magurele, Romania}
%\date{\today}

\begin{abstract}
$(\alpha,\gamma)$ and $(\alpha$,n) reaction cross sections recently measured close to the reaction thresholds are rather well described by a previously developed regional optical potential. Thus, particular features of the $\alpha$-particle optical potential at energies below the Coulomb barrier, besides parameters describing $\alpha$-particle elastic scattering at higher energies are confirmed. Additional limitations of similar statistical model calculations for minor reaction channels are shown to be most likely due to an overlooked process or critical values of statistical model parameters around closed nuclear shells.
\end{abstract}

%\keywords{}

\maketitle

The still poor knowledge of the $\alpha$-particle optical potential at astrophysically relevant energies ($E_{\alpha}\approx$ 5-15 MeV) has also motivated recent studies of $(\alpha$,n) reactions on $^{92,94}$Mo isotopes and $\alpha$-capture on $^{112}$Sn \cite{wr08} and $^{117}$Sn \cite{icd08}. They have had a key role in avoiding optical model potential (OMP) uncertainties of the statistical model calculations for reaction cross sections based on global OMPs established by analysis of $\alpha$-particle elastic scattering. Since the Coulomb barrier rules out elastic scattering measurements at lower energies of astrophysical interest, the analysis of reaction cross sections within this energy range is the only way to validate the related accuracy of an  $\alpha$-particle optical potential. Thus Rapp et al. \cite{wr08} performed useful comparisons of statistical model calculations with different optical potentials, including a recent regional parameter set \cite{ma03}. Their results emphasized either a data  overestimation by a factor of 2 or an underestimation by a similar factor. On the other hand, additional limitations of the OMP parameters, the more recent being again that of Ref. \cite{ma03}, were found below the Coulomb barrier and considered typical for the global OM parameterizations available. We provide an additional account of the new measured data by means of a recent optical potential  \cite{ma08}.

In the first place, we think it is important to emphasize the particular precondition and aims of our former OMP \cite{ma03}, which has recently been used by Refs. \cite{wr08,icd08}. Firstly we had focused on two main questions that are still open, namely the OMP parameter sets obtained from $\alpha$-particle elastic scattering at high energies ($E_{\alpha}>$ 80 MeV) which describe neither the lower--energy ($<$40 MeV) elastic scattering nor complete fusion data, and the statistical $\alpha$-particle emission that is underestimated by the OMPs that account for elastic scattering on the ground--state (g.s.) nuclei. Thus, as stated from the beginning in Ref. \cite{ma03}, we started with the analysis of the low--energy $\alpha$-particle elastic scattering alone, in order to understand at a later stage the failure of related OMP to describe reaction data. 

At the same time we noted that neither the available experimental $\alpha$-induced nor the $(n,\alpha)$ reaction cross sections were taken into account in order to avoid additional difficulties because of the remaining parameters needed in the statistical models \cite{pd02}. The differences expected for the $\alpha$-particles in the incoming and outgoing channels (\cite{ma06} and Refs. therein) have no longer played any role. A further step concerning the $\alpha$-induced reactions below the Coulomb barrier B has just been carried out \cite{ma08} while the eventual difference of $\alpha$-particle potentials in the entrance/exit channels has yet to be understood. Nevertheless, it is the potential of Ref. \cite{ma08} that should be considered within the new measured data \cite{wr08,icd08} analysis.

Basically, a regional parameter set based entirely on $\alpha$-particle  elastic--scattering analysis for mass $A\sim$100 nuclei at energies below 32 MeV  \cite{ma03} was extended to $A\sim$50--120 nuclei and energies from $\sim$13 to 50 MeV. The correlation of this phenomenological OMP to a former semi--microscopic potential based on the Double Folding Model decreases the number of free parameters considered at the same time. Then, an ultimate assessment of $(\alpha,\gamma)$, $(\alpha$,n) and $(\alpha$,p) reaction cross sections concerned target nuclei from $^{45}$Sc to $^{118}$Sn and incident energies below 12 MeV. The former diffuseness of the real part of optical potential as well as the surface imaginary--potential depth have been found responsible for the actual difficulties in the description of the reaction data below B, and have been modified in order to obtain a regional optical potential (ROP) which describes equally well both the low energy elastic--scattering and induced--reaction data of $\alpha$-particles. In order to point out the corresponding changes, the new data of Rapp et al. \cite{wr08} are compared with the results of statistical model calculations \cite{ma08,ma06} using both the parameter set suitable for the lowest energies in Table 3 \cite{ma08}, and the parameter values provided by the only elastic--scattering analysis above $B$ (Fig. 1, left side). The $\alpha$-particle total reaction cross sections of the latter case is compared at the same time with the results of previous ROP form \cite{ma03}, involved in \cite{wr08,icd08}, while the similar final values are shown together with the cross sections for all main reaction channels (Fig. 1, right side).

\begin{figure}
\resizebox{1.0\columnwidth}{!}{\includegraphics{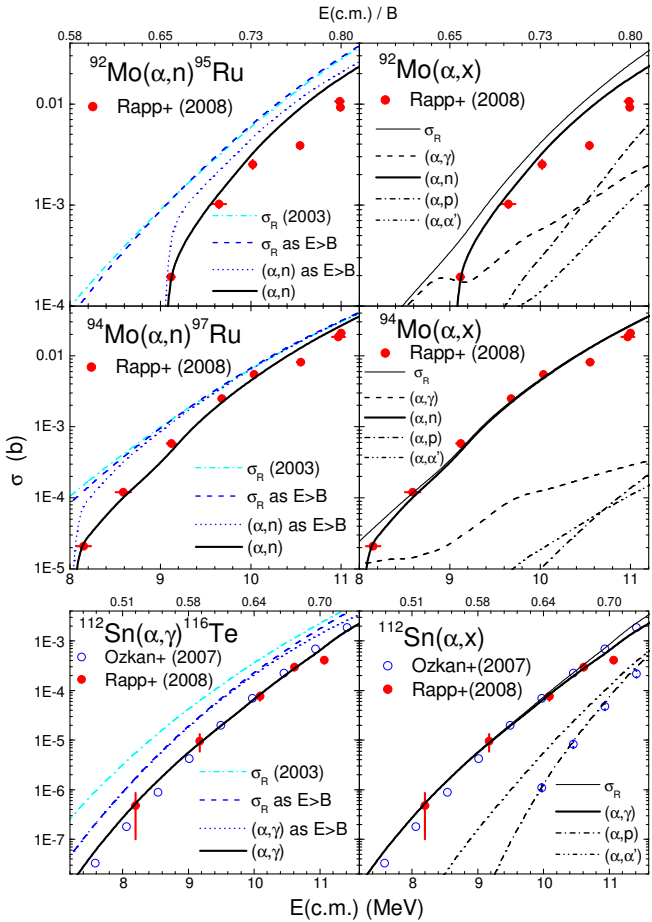}}
\caption{\label{Fig1}(Color online) Comparison of measured $(\alpha,n)$ and $(\alpha,\gamma)$ reaction cross sections for $^{92,94}$Mo and $^{112}$Sn \cite{wr08} target nuclei, respectively, and calculated values using (left) ROP parameters established by elastic--scattering analysis alone (dotted curves), and final ROP \cite{ma08} (solid); total $\alpha$-reaction cross sections provided by a former ROP \cite{ma03} (dash--dotted) and  parameters established at $E>B$ \cite{ma08} (dashed) are shown to prove the weight of the reactions under analysis. (right) Calculated cross sections for all $\alpha$-induced reactions on the same  nuclei using the final ROP \cite{ma08}.}
\end{figure}

\begin{figure}
\resizebox{1.0\columnwidth}{!}{\includegraphics{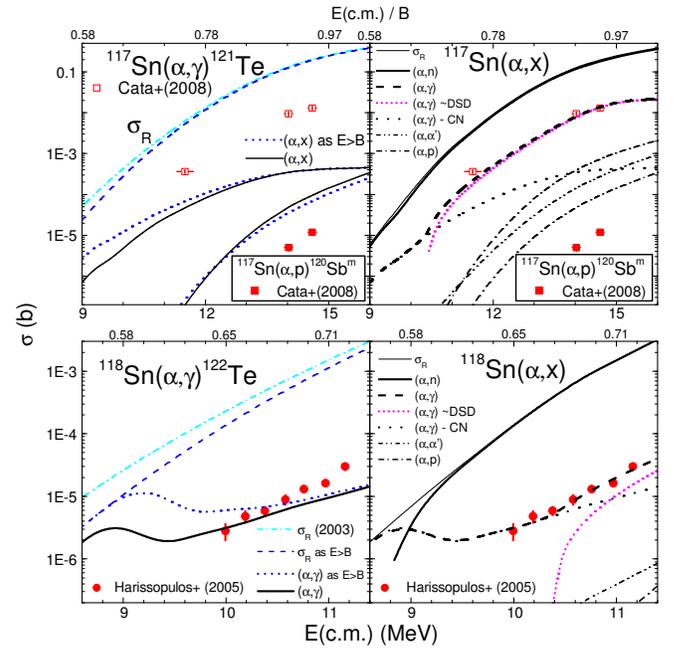}}
\caption{\label{Fig2}(Color online) The same as in Fig. 1 but for target nuclei $^{117,118}$Sn \cite{icd08,sh05}. The upper and lower dotted, solid (left) and dash-dotted curves (right) correspond to the $^{121}$Te residual nucleus ground and isomeric states, respectively.}
\end{figure}

The overall good agreement shown in Fig. 1 between the measured and calculated cross sections for major $\alpha$-induced reaction channels provides thus a trustworthy confirmation of the related $\alpha$-particle OMP \cite{ma08} well below the Coulomb barrier. The only underestimation of the new data around the incident energy of 11 MeV could arise because of a nuclear process not taken into account, as e.g. a giant quadrupole resonance which is located in $^{92}$Mo at 14.1 MeV with a width of 4.5 MeV \cite{gd88}. However it does not seem to have a major effect on the OMP validation, provided that the $(\alpha$,n) reaction cross section is suitable described within the more critical energy range just above the threshold. The reproduction of the energy dependence of new data has been more significant for the reaction $^{94}$Mo$(\alpha$,n)$^{97}$Ru, which proved to be a real challenge for the statistical model calculations \cite{wr08}. A similar case emerges for the reaction  $^{112}$Sn$(\alpha,\gamma)^{116}$Te, where a larger negative $Q$-value by 4 MeV for the $(\alpha$,n) reaction pushes its threshold beyond the concerned energy range. Since these energies are lower with respect to B than for Mo isotopes, there are larger differences between the experimental cross sections and calculations using an OMP established at $E_{\alpha}>B$ (\cite{wr08} and left side of Fig. 1). However, these cross sections are even better described by the OMP of Ref. \cite{ma08} than reaction data \cite{no07} that were part of the procedure to derive this potential.

\begingroup
\squeezetable
\begin{table*}%[H] add [H] placement to break table across pages
\caption{\label{densp} Low-lying levels number $N_d$ up to excitation energy $E_d$ \protect\cite{ensdf} used in cross-section calculations, and the levels and $s$-wave neutron-resonance spacings $D_0^{exp}$ in the energy range $\Delta$E above the separation energy $S$, for the target-nucleus g.s. spin $I_0$, fitted in order to obtain the BSFG level-density parameter {\it a} and g.s. shift $\Delta$ (for a spin cut-off factor calculated with a variable moment of inertia between half and 75\% of the rigid-body value, from g.s. to S, and reduced radius r$_0$=1.25 fm).}
\begin{ruledtabular}
\begin{tabular}{cccccccccc} 
Nucleus   &$N_d$&$E_d$& \multicolumn{5}{c}
                     {Fitted level and resonance data}& $a$ & $\Delta$ \\
\cline{4-8}
           &  &      &$N_d$&$E_d$&$S+\frac{\Delta E}{2}$&
                                     $I_0$&$D_0^{exp}$ \\ 
           &  & (MeV)&   & (MeV)& (MeV)&  &(keV)&(MeV$^{-1}$) & (MeV) \\ 
\hline
$^{112}$Sn&21&2.989&21&2.99&      &   &            & 13.85& 1.34 \\
$^{117}$Sn&21&1.710&21&1.71&11.059& 0 & 0.38(13)   & 13.80& 0.12 \\
$^{118}$Sn&38&3.057&38&3.06& 9.326&1/2& 0.055(5)   & 13.55& 1.10 \\
$^{115}$Sb&11&1.755&11&1.76&      & &              & 14.20& 0.45 \\
$^{120}$Sb&21&0.448&21&0.45&      & &              & 13.75&-1.35 \\
$^{121}$Sb&23&1.659&23&1.66&      & &              & 14.00& 0.10 \\
$^{115}$Te& 3&0.280& 3&0.28&      & &              & 14.40&-0.45 \\
$^{116}$Te&11&2.119&11&2.12&      & &              & 14.00& 0.80 \\
$^{120}$Te&20&2.461&20&2.46&      & &              & 14.00& 0.87 \\
$^{121}$Te&20&0.830&29&1.02&      & &              & 14.30&-0.72 \\
$^{120}$Te&25&2.594&25&2.59&      & &              & 14.20& 0.94 \\
\end{tabular}	 
\end{ruledtabular}
\end{table*}
\endgroup

A quite different case is that of the $(\alpha,\gamma)$ and  $(\alpha$,p) reactions on $^{117}$Sn \cite{icd08} within an incident energy range that is closer although still below B.  The related $(\alpha$,n) reaction $Q$-value being almost half of that for the $^{112}$Sn target nucleus, this reaction channel is by far the strongest at the concerned energies. Under these circumstances the differences between the reaction cross sections calculated by using the $\alpha$-particle OMPs based on the elastic--scattering analysis above the Coulomb barrier and those taking into account the reaction data as well, e.g. \cite{ma08}, are already rather small (Fig. 2, left side). However, the disagreement between the new measured data and the calculated values goes up by over an order of magnitude for both minor reaction channels. We compare it with the case of  $^{118}$Sn$(\alpha,\gamma)^{122}$Te reaction \cite{sh05}, already considered within the reaction data analysis of Ref. \cite{ma08}. The minor character of the radiative capture channel is similar, in spite of an $(\alpha$,n) reaction negative $Q$-value with $\sim$2 MeV higher. However, the above--mentioned disagreement has mainly concerned in this case the slope of the $(\alpha,\gamma)$ excitation function (Fig. 2). In order to understand the large divergence of the measured and calculated cross sections of $\alpha$-induced reactions on $^{117}$Sn, we further examined the statistical model parameters formerly adopted \cite{ma08}.

Actually, from the very beginning we strived for a better knowledge of the neutron OMP and $\gamma$-ray strength functions focusing on the analysis of neutron total cross sections for all Sn and Te stable isotopes as well as on the neutron capture on the same target nuclei \cite{va07}. Consequently, we found that the global and local neutron OMPs of Koning and Delaroche \cite{ajk03} describe well the more recent data of the total neutron cross sections for Sn isotopes, but in the limit of $\sim$15\% underestimation for Te isotopes. In order to avoid this uncertainty, the local OMP parameter set for the isotope $^{128}$Te has been adopted together with the use of Fermi--energy global values \cite{ajk03} for each Te isotope. A suitable description of the corresponding neutron resonance data \cite{ripl2} has also been checked. Next, these neutron OMPs were involved within the neutron capture analysis for all stable isotopes of Sn and Te, for the neutron energies up to 3 MeV, at the same time with recently obtained \cite{va02} nuclear level density parameters. Actually the systematical analysis of this  neutron--capture data basis was carried out in order to adopt a suitable normalization of accurate $\gamma$-ray strength functions \cite{va07} by means of independent experimental information. Nevertheless, the most important model parameters have been related to the nuclear level density. The back--shifted Fermi gas (BSFG) formula has been used for the excitation energies below the neutron--separation energy, with the parameters $a$ and $\Delta$ obtained by a fit of the recent experimental low--lying discrete levels \cite{ensdf} and $s$-wave nucleon resonance spacings $D_0$ \cite{ripl2}. For nuclei without resonance data, the smooth--curve method was adopted \cite{chj77} for the $a$ parameter of the even--even, odd--odd, and odd--mass nuclei, leading to $a$-values that were next kept fixed during the fit of low--lying discrete levels. The eventually updated parameter values, due to the structure data published in the meantime, are given in Table I together with the fitted data. 

\begin{figure}[b]
\resizebox{1.0\columnwidth}{!}{\includegraphics{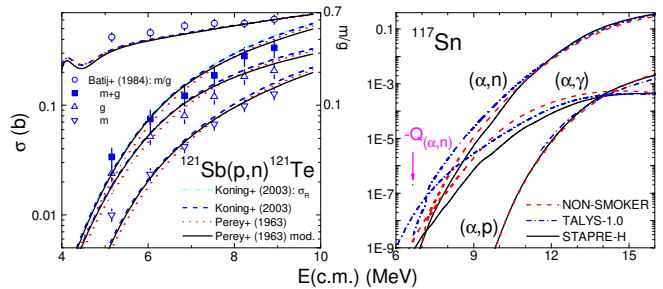}}%
\caption{\label{Fig3}(Color online) Comparison of (left) measured \cite{vcb83} (p,n) reaction cross sections and corresponding isomeric cross-section ratio for $^{120}$Sb target nucleus, and calculated values by using proton global OMPs of Refs. \cite{ajk03} (dashed curves  and dash--dotted curve for proton reaction cross section) and \cite{cmp76} (dotted), and the modified potential (solid). (right) $(\alpha,\gamma)$, $(\alpha$,n) and $(\alpha$,p) reaction cross sections calculated by using the codes NON-SMOKER \cite{tr01} (dashed) and TALYS-1.0 \cite{talys} (dash--dotted), and present local parameter set \cite{ma08} (solid).}
\end{figure}

Having obtained an overestimation of the measured $^{117}$Sn$(\alpha$,p)$^{120}$Sb$^m$ reaction cross sections, we decided to focus on the proton OMP. The only independent experimental data we have been able to use in this respect are the $^{121}$Sb(p,n)$^{121}$Te$^{g,m}$ reaction cross sections \cite{vcb83} for the production of both ground (1/2$^+$, 19.16 d) and isomeric (11/2$^-$, 154 d) states of the same residual nucleus as for the $(\alpha,\gamma)$ reaction on $^{117}$Sn. Because this reaction cross section is equal with the almost whole proton reaction cross section, its model calculation sensitivity to the proton OMP is the largest. Thus, we found that the use of the global proton OMP \cite{ajk03} corresponds to an increase of the (p,n) reaction cross sections with the incident energy a bit larger than for the measured data. The adoption of the global parameter set of Perey and Perey \cite{cmp76} with a decreased depth $W_D$=9.5 MeV of the OMP surface--imaginary part proved to be an easy way to overcome this minor drawback (Fig. 3). On the other hand, this analysis has validated, by means of the related isomeric cross-section ratio, the $\gamma$-decay scheme of the $^{121}$Te nucleus and the level density angular--momentum distribution given by the assumed moment of inertia. The slight underestimation of the measured isomeric cross-section ratio only at lower incident energies could be related to uncertainties of the $^{121}$Te residual nucleus level scheme.%, only three levels up to excitation energy of 830 keV being known to feed the isomeric state.% However, the calculated value of this ratio does not really depend on the proton OMP change (Fig. 3), while the corresponding decrease of the proton reaction cross section is only $\sim$10\%.

We have also compared the $\alpha$-induced reaction cross sections, calculated in the present work for the target nucleus $^{117}$Sn by using a local parameter set, with results of the standard model calculations performed with the well-known computer codes NON-SMOKER \cite{tr01} and TALYS-1.0 \cite{talys}. Larger differences between these results have been found only at lower energies, around and within 2--3 MeV above the $(\alpha$,n) reaction threshold (Fig. 4). However, even under such conditions, a common feature has been the energy within less than 1 MeV above this threshold, where the compound nucleus may deexcite rather equally through $\gamma$-ray and neutron emissions. At higher incident energies the neutron emission is prevailing by orders of magnitude so that only a strong effect could notably increase the $(\alpha,\gamma)$ reaction cross sections.

Finally we conclude that the disagreement of the measured and calculated $^{117}$Sn$(\alpha,\gamma)^{121}$Te reaction cross sections may be due to an overlooked process which has not been taken into account, such as the non-statistical $\gamma$-emission from the composite nucleus. The direct radiative capture in $\alpha$-induced reactions was formerly pointed out at $E_{\alpha}\approx$ 10--11 MeV \cite{kr80}, and in the mass range $A$=61--181 at incident energies of 11--27 MeV \cite{jab96}. Moreover, favorable conditions of non-negligible direct capture contribution, as both formation of the compound nuclei at low excitation energy and related low level density also  for closed shell nuclei (e.g., \cite{tr07}, are well matched for  $\alpha$-particles incident on $^{117}$Sn. It is particularly considered that non-equilibrium $\alpha$-particle captures involve $\alpha$-particles of 10--12 MeV \cite{tr08}, which are consistent with the energy range discussed in this work. However, the direct cross section is very sensitive to the predicted properties of the final states and different microscopic models yield vastly different results, while actual attempts concern the employment of averaged properties for direct captures as well \cite{tr07}. Therefore, a straight way has been adopted to presume its size for the present case, namely the use of the direct and semi--direct (DSD) formula for fast neutron capture as given by Eq. (8) of Ref. \cite{dgg82}. We have only replaced the neutron separation energy in this formula by the same quantity for the $\alpha$-particle, and a factor (1-0.67*$B/E_{cm}$) was added for taking into account the effective Coulomb barrier. Finally we used a normalization to the measured $(\alpha,\gamma)$ reaction cross section at $E_{\alpha}\approx$15 MeV, similarly to that at 14--MeV cross section data for neutrons \cite{dgg82}, with reasonable results (Fig. 2). 

Consideration of this like--DSD cross section for decreasing the $\alpha$-particle total reaction cross section is obviously followed by no real change of the $(\alpha$,p) reaction cross sections. The large overestimation of the measured data could be reduced by, e.g., the decrease of the negative backshift $\Delta$ of the residual nucleus $^{120}$Sb (Table I) by 200 keV, a similar decrease of the positive $\Delta$ for the $^{120}$Te nucleus populated by neutron emission, and an increase of the $^{120}$Te level density parameter $a$ by 12.5\%. Thus it is possible to obtain $(\alpha$,p) reaction cross sections lower by $\sim$27\%, 22\% and 69\%, respectively, as well as a smaller variance of the calculated and measured data. It is thus shown that the $\alpha$-particle optical potential is not at the origin of these problems, while the new data for $^{92,94}$Mo and $^{112}$Sn nuclei support the recent potential \cite{ma08}.

\bigskip
%\begin{acknowledgments}
Work supported by CNCSIS Contract No. 149/2007. 
%\end{acknowledgments}

\end{document}